\documentclass[conference]{IEEEtran}
\IEEEoverridecommandlockouts
\usepackage{cite}
\usepackage{amsmath,amssymb,amsfonts}
\usepackage{algorithmic}
\usepackage{graphicx}
\usepackage{textcomp}
\usepackage{xcolor}
\def\BibTeX{{\rm B\kern-.05em{\sc i\kern-.025em b}\kern-.08em
    T\kern-.1667em\lower.7ex\hbox{E}\kern-.125emX}}
\usepackage{multirow}
\usepackage{gensymb}
\usepackage{subcaption}
\usepackage{boldline}

\begin{document}

\title{Efficient Blind-Spot Neural Network Architecture for Image Denoising}

\author{
	\IEEEauthorblockN{ David Honz\'atko, Siavash A. Bigdeli, Engin T\"{u}retken, and L. Andrea Dunbar}
	\IEEEauthorblockA{
		\textit{CSEM}, 
		Neuch\^atel, Switzerland \\
	}
}


\maketitle


\begin{abstract}
Image denoising is an essential tool in computational photography.
Standard denoising techniques, which use deep neural networks at their core, require pairs of clean and noisy images for its training.
If we do not possess the clean samples, we can use blind-spot neural network architectures, which estimate the pixel value based on the neighbouring pixels only.
These networks thus allow training on noisy images directly, as they by-design avoid trivial solutions. 
Nowadays, the blind-spot is mostly achieved using shifted convolutions or serialization. 
We propose a novel fully convolutional network architecture that uses dilations to achieve the blind-spot property.
Our network improves the performance over the prior work and achieves state-of-the-art results on established datasets.
\end{abstract}


\section{Introduction}
Image denoising is a fundamental challenges in computational photography. 
With recent advances in deep neural networks, the state-of-the-art has seen a significant boost in term of reconstruction quality~\cite{zhang2017beyond}. 
These results are, however, conditioned on the availability of the pairs of noisy and clean, ground-truth, images at training time.

Lehtinen et al.~\cite{lehtinen2018noise2noise} proposed the \emph{Noise2Noise} approach that removed the need for clean images by formulating the problem as a mapping between two noisy images of the same scene. 
Assuming zero-mean noise, the network is forced to output the expected value of the pixels, which is the clean image.
However, the requirement for having two noisy instances of the same scene is quite impractical~\cite{plotz2017benchmarking}.

Assuming un-correlated noise, Krull et al.~\cite{krull2019noise2void} created a training scheme \emph{Noise2Void} that allows the network to be trained only with one noisy image per scene.
It uses the concept of \emph{blind-spot networks}, where the network cannot read the noisy pixel it predicts i.e. its receptive field does not contain the central pixel. 
This way, the model learns to predict the pixel independent of its actual noisy value.
The blind-spot property is achieved by masking the input and careful selection of the output pixels. However, the significant reduction in the pixels contributing to the loss function makes the training inefficient.




Laine et al.~\cite{laine2019high} solved this inefficiency by using shifted convolutions \cite{van2016conditional}. Their network, which we denote here as \emph{BlindSpot4D}, shifts the feature-maps so that the receptive field grows in one direction only.
The blind-spot property is achieved by feeding the network with all $90\degree$ rotations of the image and by joining the outputs using several 1x1 convolutional layers.
A limiting factor is that, by construction, it assumes rotation invariance of the kernels, which makes the network learn from an unrealistic prior in terms of the natural arrangement of the scene.
Moreover, due to the overlapped receptive fields of individual passes, the input pixels are processed twice, resulting in high computational effort.

\begin{figure}[t]
	\centerline{\includegraphics[width=0.95\columnwidth]{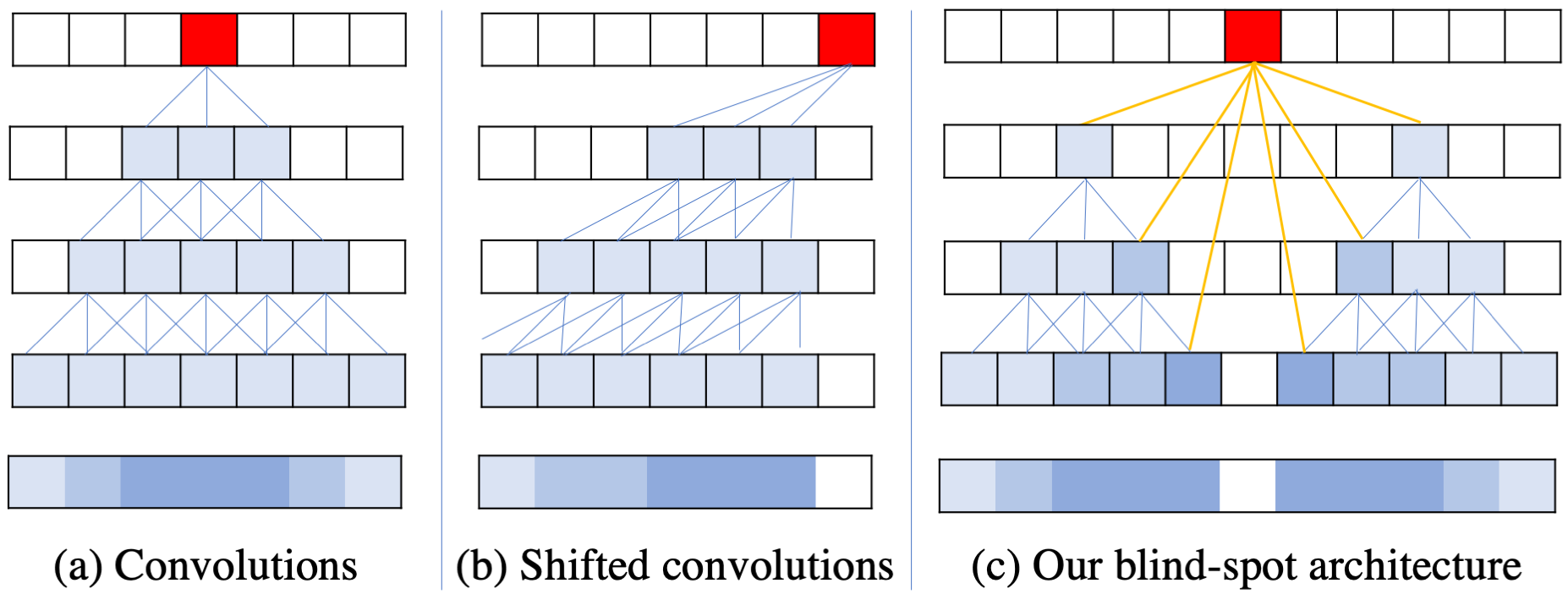}}
	\caption{Receptive-field of different network architectures.}
	\label{fig:netrecept}
    \vspace{-8pt}
\end{figure}


We propose an alternative network architecture that guarantees the blind-spot property by design in a single pass and tackles the previously mentioned drawbacks -- rotation invariance and redundant computations. In the following, we present our architecture and show the state-of-the-art results.


\begin{figure}[b]
	\vspace{-8pt}
	\centerline{\includegraphics[width=0.95\columnwidth]{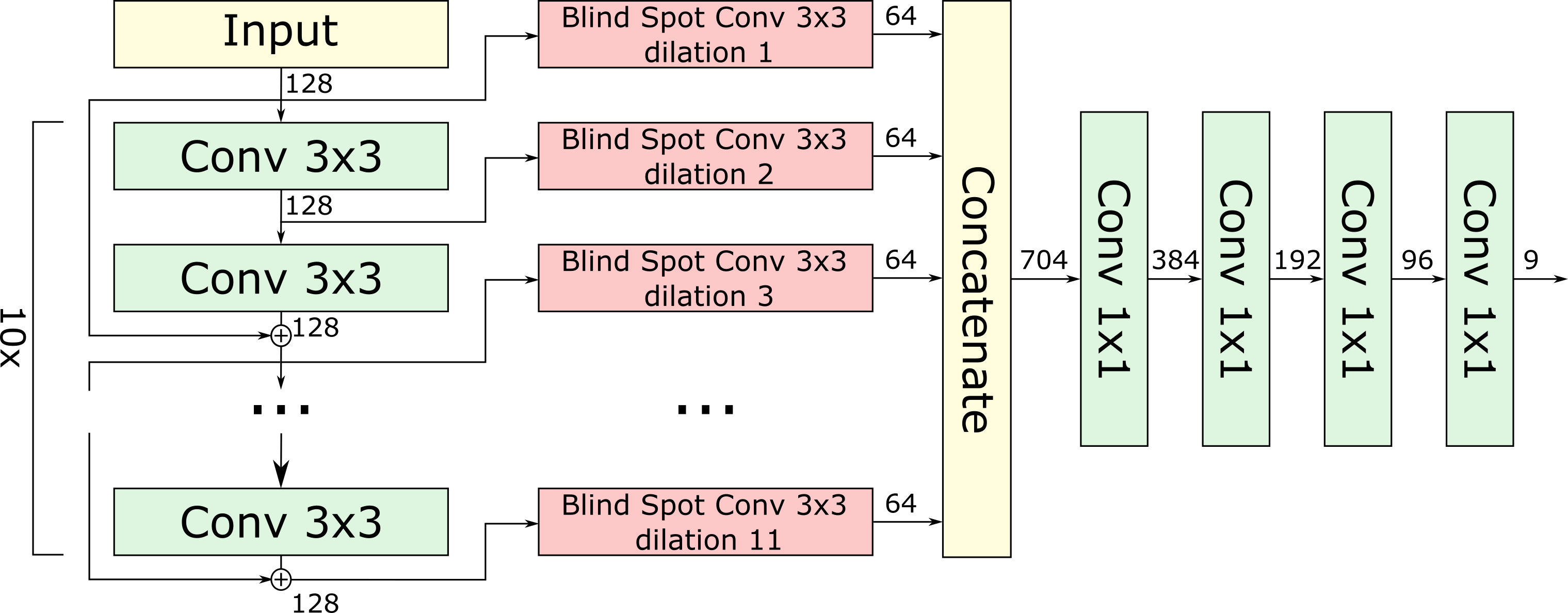}}
	\caption{Our network architecture with a depth of 10 layers.}
	\label{fig:arch}
	\vspace{-8pt}
\end{figure}

\section{Blind-spot Architecture}

The essential building block of our network is a dilated convolution with a blind-spot in its kernel. The dilation allows us to connect an arbitrary number of preceding convolutional layers while maintaining the blind-spot property of the network as depicted in Figure~\ref{fig:netrecept}. 

We show the complete architecture in Figure~\ref{fig:arch}. 
The forward stream comprises conventional convolutional layers, where we use residual connections between each second layer to improve training stability.
The result of each convolutional layer is then fed to a blind-spot convolutional layer dilated by one plus half of the size of the receptive field of the respective input layer.
Finally, the outputs of these dilated convolutions are concatenated and passed through several $1\times1$ convolutions to predict the results.


\begin{figure}[b]
    \vspace{-15pt}
	\centering
	\begin{subfigure}{.425\columnwidth}
		\centering
		\includegraphics[width=\textwidth]{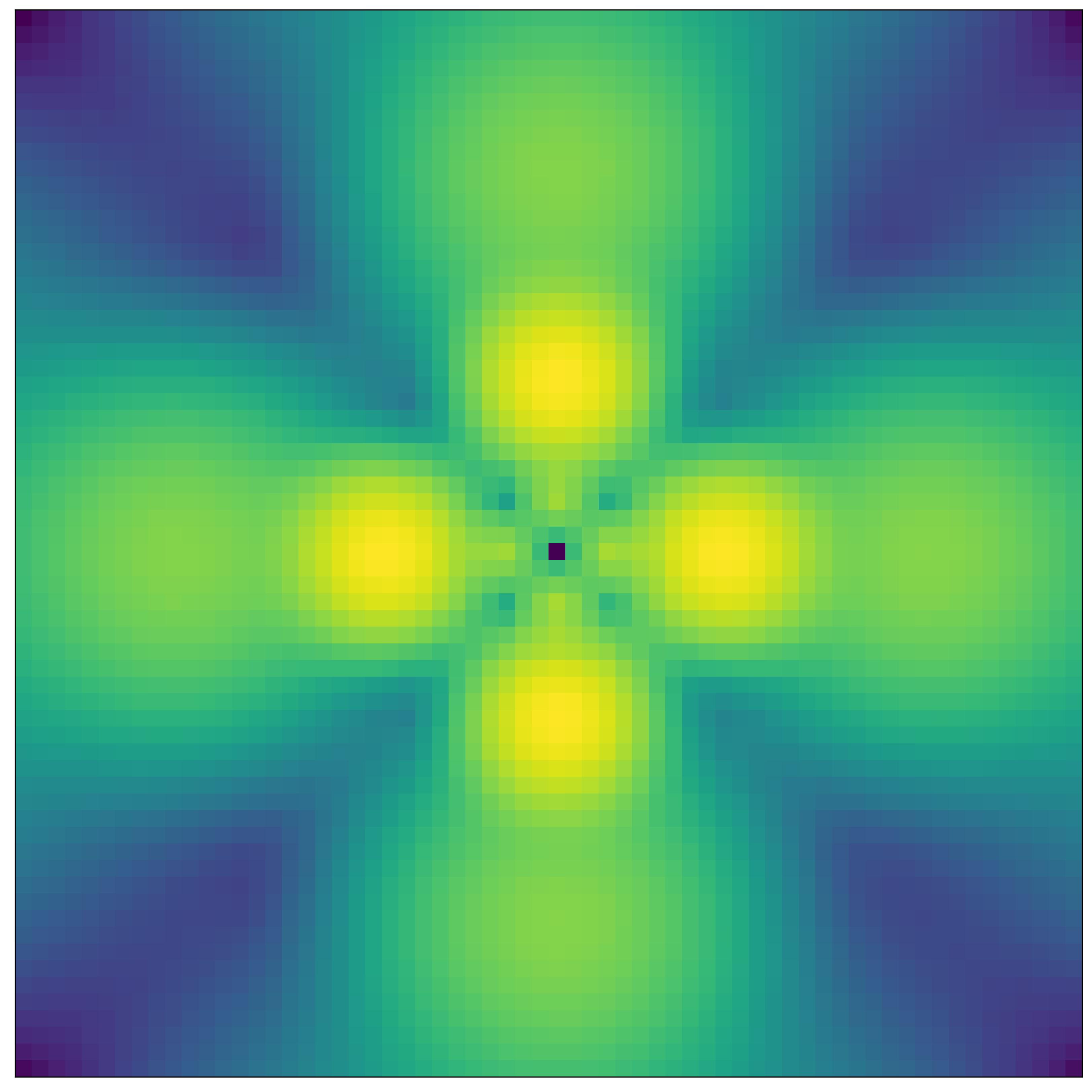}
		\caption{BlindSpot4D ($570\times570$)}
		\label{fig:receptivefieldcomparison:bs4d}
	\end{subfigure}%
	\begin{subfigure}{.425\columnwidth}
		\centering
		\includegraphics[width=\textwidth]{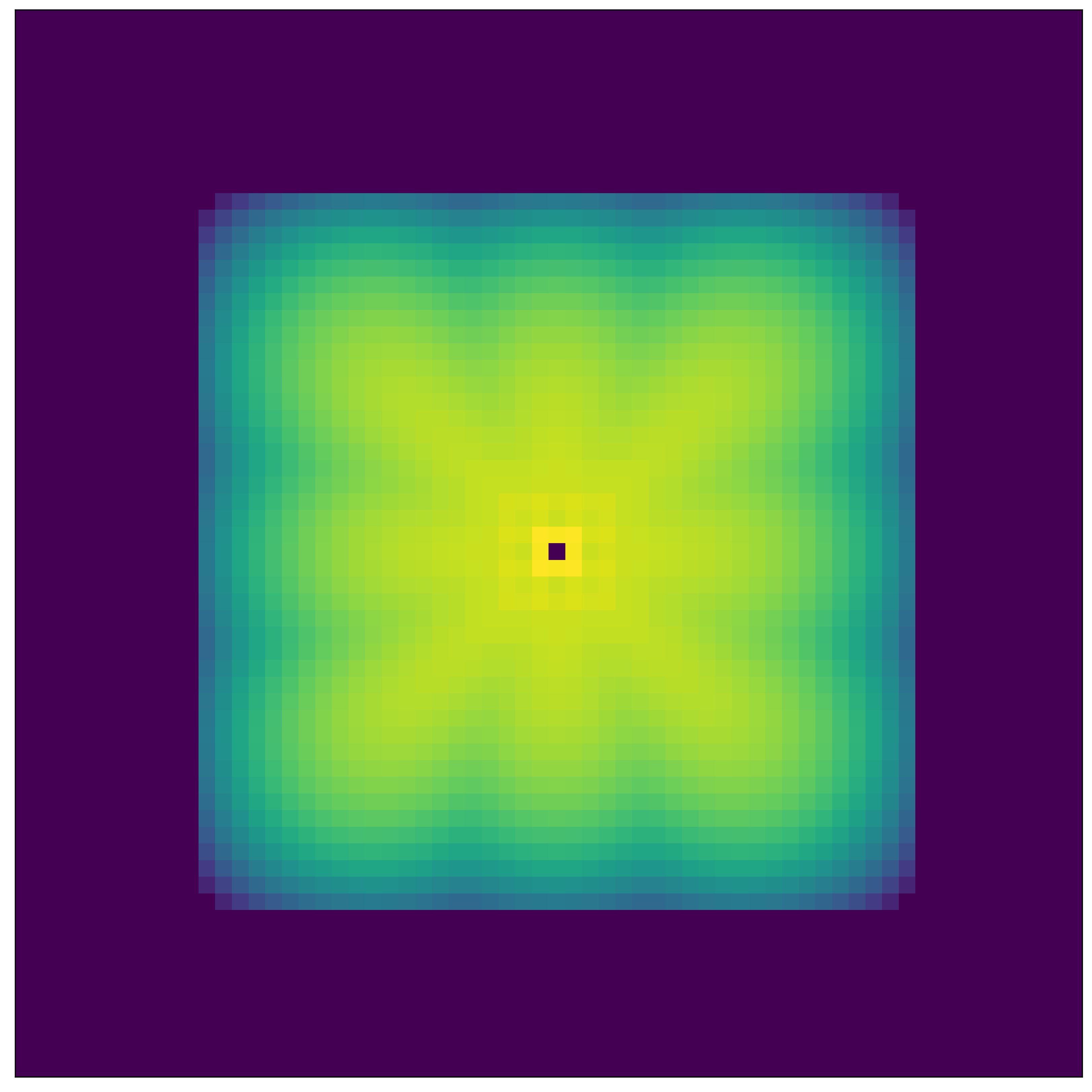}
		\caption{Ours ($43\times43$)}
		\label{fig:receptivefieldcomparison:ours}
	\end{subfigure}
	\caption{Visualisation of the receptive field in log-scale. Images are cropped to $64\times64$ and normalized for visual clarity.}
	\label{fig:receptivefieldcomparison}
    \vspace{-8pt}
\end{figure}

The blind-spot convolutions with different dilations allow the network to remain focused on the immediate neighbourhood of each output pixel.
They also ensure integrity of the receptive fields for pixels closer to image borders.
Figure~\ref{fig:receptivefieldcomparison} depicts the receptive field of our network compared to the BlindSpot4D architecture. 
These are the responses of the networks to the Dirac function, showing the number of visits the network has made for each pixel before inferring the results.
The receptive field of BlindSpot4D shows low coverage of the immediate neighbourhood of the central pixel and uneven distribution in diagonal directions.
Conversely, our network concentrates more on the close neighbourhood of the centre pixel and covers all the directions equally.
This is an essential feature for image denoising, where the more distant areas in the image are usually less relevant.

\section{Experiments}

Following the framework devised by Laine et al.~\cite{laine2019high}, we test our architecture on several image datasets and noise distributions.
For a fair comparison, we used the same prediction scheme, objective function, and dataset to train our network.
The output of the network is the mean and covariance matrix of a multivariate Gaussian distribution.
The final prediction is estimated as a product of the output and the noise distributions, centred at the noisy input value.
The results in Table~\ref{tbl:psnr} show that our model achieves state-of-the-art results in terms of peak signal-to-noise ratio (PSNR).
These results are consistent in all datasets and noise distributions.

\begin{table}[t]
	\caption{Image denoising performance comparison using PSNR scores for different noise distributions and datasets.}
	\begin{center}
		\begin{tabular}{l c c c c c}
			\hlineB{3}

			\textbf{Method} & $\sigma$ known? & \textbf{KODAK} & \textbf{BSDS300} & \textbf{Set14} & \textbf{Avg.} \\
			\hline
			\multicolumn{6}{c}{Gaussian $\sigma=25$} \\
			\hline
			Noise2Noise   & no & 32.45 & 31.07 & 31.23 & 31.58 \\
			BlindSport4D & yes & 32.45 & 31.03 & 31.25 & 31.57\\
			CBM3D           & yes & 31.82 & 30.40 & 30.66 & 30.96\\
			Ours			   & yes & 32.45 & 31.02 & 31.25 & 31.57 \\
			
			\hlineB{3}
			\multicolumn{6}{c}{Gaussian $\sigma \in [5,50]$} \\
			\hline
			Noise2Noise   & no & 32.57 & 31.29 & 31.26 & 31.70\\
			BlindSport4D & yes &32.47 & 31.19 & 31.21 & 31.62 \\
			CBM3D           & yes & 31.99 & 30.67 & 30.78 & 31.15 \\
			Ours			   & yes & 32.46 & 31.18 & 31.25 & 31.63 \\
			
			\hlineB{3}
			\multicolumn{6}{c}{Poisson $\lambda=30$} \\
			\hline
			Noise2Noise   & no & 31.80 & 30.39 & 30.44 & 30.88 \\
			BlindSport4D & yes & 31.65 & 30.25 & 30.29 & 30.73\\
			Ours				 & yes & 31.67 & 30.25 & 30.14 & 30.69 \\
			\hlineB{3}
		\end{tabular}
		\label{tbl:psnr}
		\vspace{-8pt}
	\end{center}
\end{table}

\begin{figure}[t]
	\centering
	\begin{subfigure}{.25\textwidth}
		\centering
		\includegraphics[width=\textwidth]{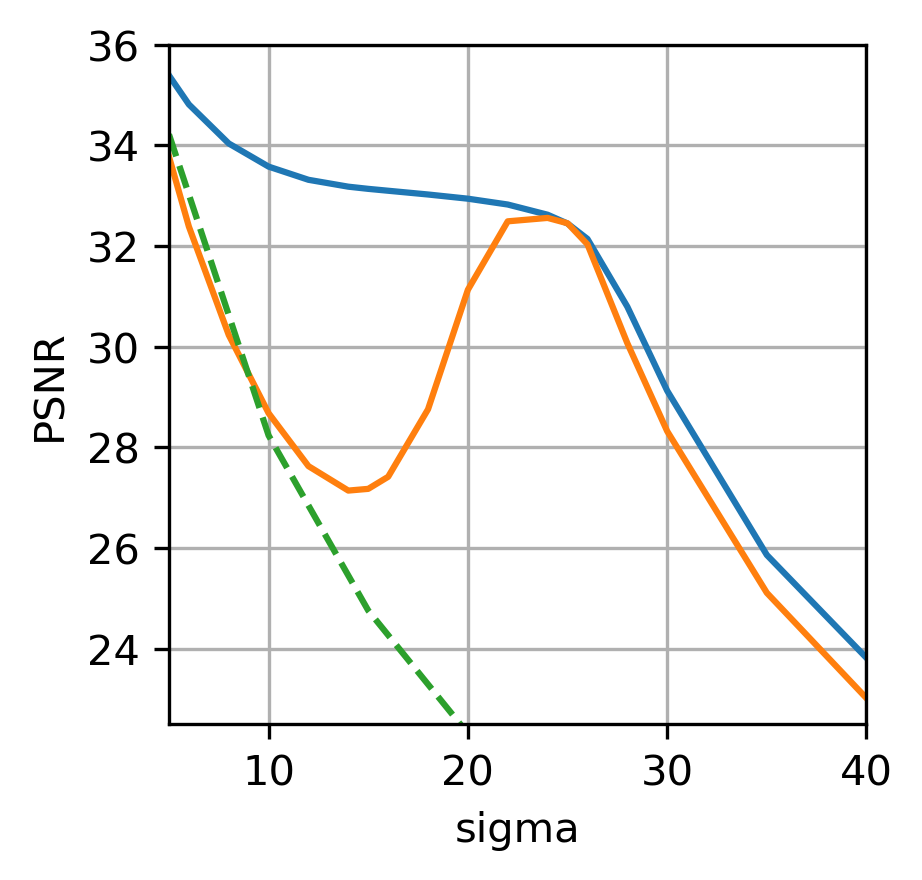}
	\end{subfigure}%
	\begin{subfigure}{.2375\textwidth}
		\centering
		\includegraphics[width=\textwidth]{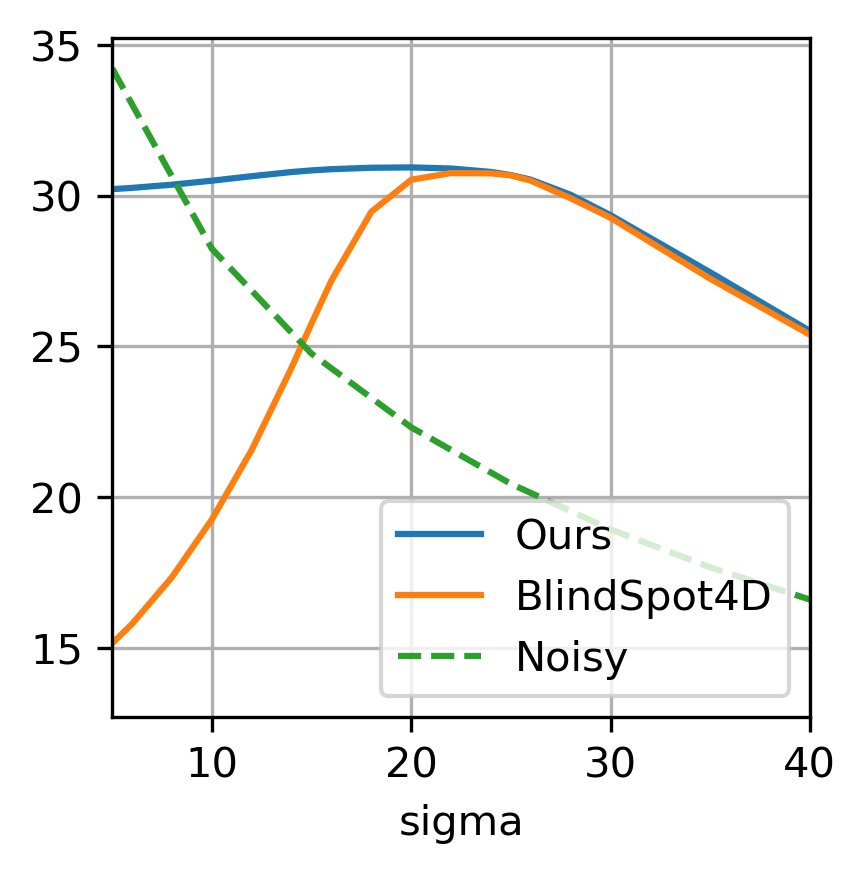}
	\end{subfigure}
	\caption{Testing the networks trained for $\sigma=25$ on different test $\sigma$. \textbf{Left}: Actual prediction. \textbf{Right}: Prediction of the mean only (not using the noisy value). }
	\label{fig:diffferentsigma}
    \vspace{-8pt}
\end{figure}

To analyse the effect of the receptive field, we also tested both blind-spot architectures for different levels of noise variance at test time.
Note that both methods were only trained for standard deviation $\sigma=25$.
We show these results in Figure~\ref{fig:diffferentsigma}, where we compare our model to BlindSpot4D.
To better evaluate the methods, we also compare the estimated means of the output distributions, independent of the corresponding noisy input values.
These results show that our method can achieve significantly better reconstruction scores (up to $6$ PSNR) for lower noise levels.
This can be explained by the shape of the receptive field, where closer pixels contribute more to the results.
For higher noise variances, this property becomes less critical as the prediction has to be made based on much larger neighbourhood.


\section{Conclusions}
We propose a new blind-spot network architecture that uses conventional training strategies.
Compared to state-of-the-art, it does not require sequential inference nor independent processing of directions.
We achieve comparable results with the state-of-the-art methods in image denoising with different noise distributions.
Our method specifically outperforms these techniques by a large margin, where the input noise variance is closer to more realistic values.

\bibliographystyle{IEEEtran}
{\bibliography{paper}}

\end{document}